# Beyond Replacement or Augmentation: How Creative Workers Reconfigure Division of Labor with Generative AI

MICHAEL F CLARKE, Independent Researcher, Canada
MICHAEL JOFFE, Google Inc., Canada

The introduction of generative AI tools such as ChatGPT into creative workplaces has sparked highly visible, but binary worker replacement and augmentation debates. This study reframes this argument by examining how creative professionals re-specify a division of labor with these tools. Through 17 ethnomethodologically informed interviews with international creative agency workers we demonstrate how roles are assigned to generative AI tools, how their contributions are modified and remediated, and how workers practically manage their outputs to reflect assumptions of internal and external stakeholders. This paper makes 3 unique contributions to CSCW: (1) we conceptualize generative AI prompting as a type of workplace situated, reflexive delegation, (2) we demonstrate that workers must continuously configure and repair AI role boundaries to maintain workplace intelligibility and accountability; and (3) we introduce the notion of interpretive templatized trust, where workers devise strategies to adapt automated generative templates for their setting, and reinforce stakeholder trust. This contribution has implications for organizing productive human-AI work in creative and stakeholder centric environments.

## 1   INTRODUCTION

Artificial intelligence in the workplace has captured the imagination of onlookers questioning how the modern workplace will change as its capabilities broaden. This has ranged from fear, to excitement, and has prompted many questions about augmentation versus replacement. In the American Psychology Association's 2024 Work in America Survey, 41% of employees said they worry that AI will make at least part or all of their job responsibilities obsolete in the future [1]. Mckinsey & Company [18] describe a broad range of revenue gains and operating cost decreases by introducing generative AI in the workplace, ranging from Human Resources to product development and even software engineering.

Given their complexity and nuance, functions previously thought difficult to automate with AI are now a possibility. The excitement of businesses seeing an AI led path to profit through replacement and augmentation is understandable. Thus, popular workplace discourse has been dominated by the replacement vs augmentation debate.

Sam Altman, CEO of OpenAI recently said: "A decade ago, the conventional wisdom was that AI would first impact physical labor, and then cognitive labor, and then maybe someday it could do creative work. It now looks like it's going to go in the opposite order"

[33]. Elsewhere Altman suggests that "95% of what marketers use agencies, strategists, and creative professionals for today will easily, nearly instantly and at almost no cost be handled by the AI" [cf.10]. Given the novelty and industry fit of generative AI Creative work, "industries which have their origin in individual, creativity, skill and talent" [9], have received disproportionate attention in these replacement and augmentation conversations. The most commonly adopted AI use cases across all industries related to jobs (albeit rudimentary) are typically done by the creative worker (writing, personalized marketing, image and video generation) [36]. A Goldman Sachs [14] report suggests "some types of work, such as logo design, copywriting, translation, or voice-over artistry, could be displaced by free or cheap AI tools in those categories".

While augmentation and replacement dominate popular discourse, they tend to gloss over the context bound, everyday commonsense methods relied upon to make generative AI tools work in the workplace. In this case of creative workers utilizing generative AI, we describe the methods creative workers rely upon to build up, respecify and preserve division of labor with AI.

We make 3 contributions to CSCW:

First, we conceptualize prompting as reflexive delegation, where creative workers interpretively configure a division of labor. Creative workers craft AI collaborator personas by relying on typified roles tied to their unique workplace context and assumptions.

Second, we introduce the idea of boundary repair with generative AI, where workers discursively and collaboratively adjust AI's job in reaction to disrepair in trust, mutual intelligibility and quality.

Third, we put forward the concept of templatizing trust that helps teams practically account for and manage tensions related to scale from automation. These tensions arise from client stakeholder needs related to deliverables.

## 2  RELATED WORK

Previous contributions concerning generative AI and creative work have focused on specific task types [19; 7; 39], contexts, and goals [e.g. 27]. Others explored how we might enhance and evaluate creative outputs [e.g. 2; 41; 15; 38], and creative workflows [e.g. 43].

A body of work provides us with more nuance behind generative AI worker augmentation arguments, such as how we might coexist to enhance creativity [40; 26; 21], including how AI tools might assist specific parts of the creative process, such as brainstorming [28].

Other work focused on generative AI business compatibility and integration issues. These include proposals for change management that specifically assist creativity and problem solving [35], using HCAI principles to guide workers through challenges including content creation [11] and understanding creative worker perceptions assistants [23].

Lastly, some have looked at how specific elements of generative AI need to be improved to support design, such as model transparency and explainability needs [20],

While several studies unpack issues that run deeper than simplistic claims of augmentation and replacement, they do not examine how creative workers practically organize work, and interpretively reconfigure division of labor with generative AI in practice.

Previous contributions focus primarily on designing interfaces, user intention to adopt tools and for what reasons, and other elements of user experience. They do not examine the reflexive, interpretive methods creative workers utilize to make generative AI usage and output accountable and intelligible to stakeholders. We address the praxeological gap [6] in the division of labor that unfolds with these tools introduced into the creative setting. In other words, the work to make the generative AI work. We turn to ethnomethodology to do this.

## 3   THEORY

Computer-Supported Cooperative Work (CSCW) and ethnomethodology are useful for understanding complicated work settings where multiple technologies are deployed. In our case these tools include novel generative AI tools (e.g. ChatGPT, Midjourney etc), well known office technology (e.g. PowerPoint, Google Slides, Google Docs, Zoom Video Conferencing) and a myriad of internal collaborators and external stakeholders.

Thus, we rely on ethnomethodology to help with detailed descriptions of practical, ordinary methods workers use to accomplish relevant and essential work, tied to the context of their workplace. Harper and Randall [16] describe how CSCW and ethnomethodology have proven useful for understanding AI and machine learning in the workplace: "Dealing with culture where technologies are used is the business of CSCW...the approach speaks naturally to what is done with AI and ML, not claims about the technologies" (p.130). They recommend we draw from Garfinkel [12] "showing how accounting for decisions which go into feeding data into systems to produce a 'ground truth' are accountable matters, matters worthy of careful investigation, just as are the ways that objective functions are accountable too" (p.131). In our case, members of the creative organizations seek relevance (and have methods for doing so), in pursuit of creative outcomes. In other words, CSCW and ethnomethodology are equipped to help describe the detailed relationship between "humanized" generative AI software and the form they take with for the work at hand for creative workers. We describe this by considering the way AI is prompted by workers, and how they make sense of what comes back, while paying attention to the "organizational contexts in which ML [and AI] technologies are interpreted, decisions are implemented, by whom and why" (p.132). As Martin et al [24] say, ethnomethodology is of "specific relevance to CSCW in situations where we may have relatively little to go on - conceptually and empirically - in efforts to open the 'black box' of creativity, as it unfolds in practice" (p.174).

Mair et al [22] provide us with an example to consider when examining the famous "move 37" by Google's DeepMind's AlphaGo. In this AI vs human match of the complex board game, they urge us to look beyond technical properties as technologies, which "are defined by how we involve them in our practices"(p.353). Taking a similar perspective, we suggest that as these new forms of creative workplaces develop, we recommend new creative technologies must include workplace practices in their composition: "In the case of AI technologies, those involvements are still developing and we need to see both saying and doing as interlinked practices central to their continuing elaboration" (p.353).

## 4 METHOD

Our study involved remotely interviewing 17 participants who work in international creative agencies over the course of 5 months (April 2024-August 2024). These interviews were semi-structured and ranged from approximately 20 minutes to 1 hour (and follow up deep dives, in some cases examining generative AI prompts and response). We used Zoom video conferencing and recorded the interviews for "subsequent analysis of participants' accounts of their methodological properties" [cf.6]. We subscribe to the "phenomenological" variant of ethnomethodology, using interviews as a commonsense resource for data. Insights from the interviews are clear and understandable but do not declare some generalized truth. In other words, we agree that ethnomethodology is not bound to specific "modalities of human action..nor research methods" (p.7) [18]. Remote interviews accessing home work environments (a common setting for creative agency workers) removed significant overhead in completing interviews, and increased the likelihood of more candid conversations with participants. This was important given the sensitivity of the topic.

## 5 SETTING

We interviewed creative managers and practitioners at 12 different creative agencies (ranging in size from 50 to over 75,000 employees) in the United States and Canada, from May 2024 to August 2024. Creatives in our study use generative AI tools as part of their work of creating media and advertising assets and/or "brand strategy". The quality of these outputs are judged based in part on the degree to which they are humanized on personality attributes that help the company, product, or brand better connect with end users (See Appendix for participant details).

Creative Practitioners "prompt" generative AI software (e.g. Chatbot, Text-To-Image/Video) in order to achieve a desired creative output from the model, be it text, image or other media, which can be seen as a type of natural language programming that generates media.

Study participants work on creative project based teams of client managers (roles called Engagement Managers, Account Directors) and sub-teams of creative managers and practitioners (roles called Brand Strategists, Creative Directors, Copywriters, Art Directors and Designers, Project Managers and Producers). Executive Creative Directors manage teams of creatives across several projects. The work takes place in offices designed to facilitate individual and collaborative work (e.g. client office, agency office), on virtual video calls (both from participants' homes and offices) mediated by software (e.g. Miro, Figma, Zoom, Google Meet).

## 6  PROMPTING FOR BUILDING UP ROLE DEFINITION AND DELEGATION

### 6.1  Role Delegation through Prompting

We learned that workers establish a division of labor, setting up roles and boundaries for work delegation using generative AI. Workers rely on context bound knowledge about what shape a typical worker would take, in order to be effective. Typifications [12] provide an accessible shared language that enables coordination and help accomplish mutual intelligibility to move to an effective outcome. Part of that effective outcome is to construct an adequate sidekick for the tasks at hand, in this case: doing creative advertising work for clients.

When workers create generative AI prompts, they aren't simply directing or operating a machine, they are configuring a coworker. Prompts are situated tools to a local organizational challenge. They are reflexive tools, and the shape that they take is dependent on the worker collaborating with them, and the situational unfolding of the task at hand. For Garfinkel [12] they are developed in order to be recognized as a competent organizational product.

The challenge is making generative AI intelligible and useful for the work at hand. Each generative AI prompt is reflexively constructed to help with the work in that instance, and draws on background expectancies and typified roles.

Here we see how specific attributes about what makes a "good" assistant are an example of drawing on language as an indexical tool tied to the context of the workplace, and are essential for defining a division of labor that contains AI as a collaborator. These descriptions of generative AI as a worker, are indexically bound to the local workplace setting, and are occasioned with "a definite sense and reference according to the place in which the utterance or expression is done" [13] (p.22)

Workers are uniquely positioned to interpretively select attributes like, age, domain expertise, and more abstract concepts like style, which are not preordained by the tools themselves, and are critical to ChatGPT and other generativeAI tools when they are "hired".

Art Director 1, who is delegating to ChatGPT, requires that his "assistant" has specialized knowledge about a specific domain and skills for creative work. He implies that prerequisite experience type and tenure are critical to building to an effective outcome:

*"I always start by kind of baking in like the requirements ahead of time…I want you to be…a professional writer in the pharmaceutical field...take your 35 year old"*

He assumes that by transforming the assistant into a 35 year old, professional pharmaceutical writer, it can produce adequate creative work for the client. He "hires" this talent, and places them in a role that complements his skill sets. He relies on his role and workplace specific knowledge to select this ideal typical profile for the specific work, and the specific client. The client trusts him to make these decisions. Tenure and professional domain are not determined by AI, nor are they a predetermined directive from management or client, and this is reasoned through by him in the flow of work. These occasioned concepts (35 year old pharma writer) are indexically tied to the work skills, experience and work setting. They make sense and are made significant only because they invoke everyday knowledge of competence and tone to produce good creative work.

This is extended to style by Art Director 1, an assumed important attribute for doing good work:

*"the character that I want it to be and I also prime it with…any literature you give…you give a writer anyways like, here's a Pdf of like…the writing style…I always make sure I feed it like..I always approach every conversation with ChatGPT like I'm building a GPT...it's conversational."*

Style assumptions are applied in different ways, depending on the role of the worker. Art Director 2 delegates image generation for a physical design installation:

*"like there is no like a like an official handbook, except, you know, some rules…but you know some people just tend to describe spaces or imagery in a in a better way, I would say…like my 3D background helps me a lot…like creating something wavy, abstract, cool architecture thing I would ust type in. Yeah, give me a Zaha [Hadid]…some people don't really know…they type and like, give me a wavy building…get a totally different result"*

The invocation of Zaha Hadid, a famous architect and artist, shows the designer drawing on what Sacks [32] calls a membership categorization device, a method for navigating culture, in this case workplace culture. We identify people and concepts as types, which are relied upon as a resource in dealing with collaborators. They are sensemaking tools used by analysts and workers alike, and applied to articulate the practical reasoning these workers are dependent upon to understand and act in their workplace. She selects

*Beyond Replacement or Augmentation: How Creative Workers Reconfigure Division of Labor with Generative AI*

a persona that connects design sensibilities and expectations, emotional tone, and design style into the job of generative AI. It gives the AI some autonomous space through categorizing who it should be, while also creating boundaries for reasonable output.

She draws on a shared, but not articulated, cultural reference as part of the practical expertise of her role. A "typical" Zaha Hadid form of the design can be utilized by generative AI because the worker is able to invoke its professional meaning in the right place and the right time for the task. She relies upon these Membership Categorization Devices to position herself as an expert in a particular type of design, and thus is able to implicitly and effectively delegate to AI to react to that social position, and effectively respond to produce something good enough for the task at hand.

We also saw how roles are shaped to fit relevant creative work tasks. For example, some will scale down the skills and scope of responsibility of Generative AI tools. A consumer researcher/analyst working in a strategy function showed us how she relies on her previous experience as a data analyst in order to shape the Generative AI tool, and subsequently how she interacts with it, to produce a quality outcome.

"*when I'm doing data analysis on like customer support tickets and trying to get a sense of like what the main themes are. I'll be like, go line by line for this particular document. For this particular excel. Categorize all of the comments by, let's say, 5 categories. Go line by line, mark each one of the particular category, and tell me what the 5 categories are, and give me something…I treat it like a really underpaid year one management consulting analyst. And I'm being a little bit of a micromanager….I understand that, like theoretically, that as I'm interacting with it and correcting it. It'll it'll tell it. That is information that will help improve it, but that the motivation for me is just to not have incorrect information*"

She scales the AI assistant down by invoking the typical work behaviors of a junior analyst, tied to the context of both her experience of what is acceptable practice, and the task she has interpretively defined as a match with the output she desires from the tool.

The same researcher relies on practical reasoning [12] to make a model selection (Claude) and delegate it to a moodboarding task, instead of using ChatGPT, which she uses for data analysis.

"*I still most use. Gpt, 4. I do really like Claude. Because it it's a little bit less like business. He, the the the outputs, and the way of interacting is a little bit less businessy. It's a little bit more creative and a little bit more human. So I do generally enjoy output where for, so if I'm doing like storyboarding, for example, I don't do a ton of for [Company], but in my other, in in previous work I would. If I'm doing a customer storyboard. I will have Claude write a script for an 8, you know 8 panel story board. I usually like to do hand illustrations because I just like craft of it. But every now and then I'll also just like pull visuals for for that. But really the way that I interact or interact with.*"

The researcher relies on lay sociological methods to marshall non-technical and technical resources to get to a good storyboarding outcome. This includes her everyday workplace knowledge of what style of output will lead to success, which is assumed to hinge on what is reasonable tone with colleagues (designers and product managers) for the storyboarding exercise. Thus, the researcher combines her knowledge of the task, the assumption of the "tone" of the AI tool (more casual), her knowledge of what good looks like for the task, and how it will come across to internal colleagues she is guiding. This is an assumed but unarticulated attribute of the exercise, that does not rely on generative AI alone, or in a simple transactional sense.

Thus, we can see that the "delegation" task when creating a division of labor is bound up in technical and non technical know-how, and workers draw on their common sense knowledge of the workplace to achieve it. They do this interpretively through typificatory practices, relying on assumed attributes that are not preordained ahead of time, nor logic contained in the generative AI tools themselves.

The prompting practices seen in these examples are not simply design practices, they are reflexive acts for creating order in this workplace. They are methods for making sure generative AI's contributions are sufficient and recognizable by all stakeholders involved. Creative workers render scenes accountable by the interpretive work of prompting, while drawing on scenic features and typifications to modify what is considered a plausible "coworker" and division of labor.

**6.2   Repairing Boundaries and Roles**

Beyond delegation, we see how workers are faced with situations where they need to modify, or reassert the role of their generative AI collaborator. They do this in order to make it fit properly in the workplace. This is done by actively and reflexively revising its contribution and boundaries.

Revising the output of generative AI, after its particular role has been established, is a necessary part of socio-technical collaboration in a workplace that includes AI as co-worker. Workers rely on practical methods to *revise* their ultimate product, and get them to a form that's good enough to proceed with a good work product for end customers.

AI role boundary repair is practical work for maintaining social order in the workplace. This order is not preordained but embedded in the methods used by the workers themselves to proceed. They reflexively draw on local knowledge to do this. Repair work is central, not just limited to revision. It is work that restores workplace order [12]. In the case of repair, this occurs through respecifying the relationship between the worker and generative AI going forward, and where its role is best situated.

Art Director 2, who designs packaging, web and communication design recalls an instance of generating imagery, and how generative AI has changed her previous work

*Beyond Replacement or Augmentation: How Creative Workers Reconfigure Division of Labor with Generative AI*

structure. She does this through an instance of generating the imagery that sits in the backdrop of her work. She takes advantage of generative AI to revise her previous work structure, which allows her to be more productive with her new "coworker(s)".

She described a specific scenario of using Photoshop and AI to bring an idea to fruition:

*"I'm usually in Photoshop, changing colors or taking a piece from this image and a piece from that image and putting them all together to try to paint the idea, and in the past it's just been that it's just me finding images, writing, putting it together myself."*

She also describes some of the specific image quality problems she encounters when working with the tool in this new work configuration:

*"What happens with cotton flowers? For some reason, when it creates a new version of it, it liquefies them. They just turn into sludge. And I'm thinking, why is this degenerating, instead of becoming more appropriate or closer to what I want"*

She then goes on to describe how she changed her workflow to accommodate the new skills introduced:

*"And then what I actually have noticed in the last 12 months of practice is that I've been able to refine my workflow. So I save a lot of time on the little frustrating, tedious things that I used to have to do like, Oh, there's a bad shadow, or there's a rip, or there's a perforation…I can get more really great retouching down without extra weeks or extra people…that being said, it's not just click and go…It's definitely you really have to learn what to type and what not to type. And you start to learn if I say this, I'm going to get that you know error. Or if I type this, it's gonna give me that. And I'm talking right now about Photoshop, Photoshop has its whole AI integration now. And so when you start learning to talk to it you get a better result"*

The inability of the tool to produce the images required is not simply a transactional breakdown or inadequacy of the technology. Rather, generative AI is being relied upon as a collaborator, and in the process the worker is assessing its adequacy. The problem of quality produces trouble for organizational order, when it fails to line up with local work intelligibility [12]. To repair this, she is able to reason through her new work configuration, and her role in relation to AI. This work configuration is not a preordained, objective fact to be discovered, it's determined reflexively through the course of work. This is seen in her drawing on her expertise in design and interpretively applying standards of what is good enough for the task. She implicitly reasons through what the AI can do now, versus what she can do, and establishes new workplace boundaries as a result. She evaluates when to step in to reconfigure its position in relation to the contextually bound knowledge and workplace practice only she can undertake. As Garfinkel [12] suggests, disregarding this lay sociological reasoning used to create this "division of labor" takes for granted "that a member must at the outset "know" the settings in which he is to operate if his practices are to serve as measures to bring particular, located features of these settings to recognizable account" (p.8). In other words, how the creative worker is configured in relation to the generative AI, and its ability to do good work or not, is not preordained but tied to the setting. She recounts her reasoning through these issues of quality and how work is delegated or retained. This reasoning cannot be done separate from the setting.

A Copywriter does something similar for writing. He has calibrated the level of seniority of ChatGPT as a co-worker, and what tasks it can reasonably undertake. He understands his role as interpretively assessing tone for a given piece of writing, and when to invoke a prompt to achieve it.

*"I'll get stuck on a phrase, or something that I feel is either not sophisticated or too sophisticated. At which point I will either dumb it down or smarten it up, using GPT again. A lot of it is understanding how to how to like properly prompt it. And then if, if, as it's generating writing, and I'm and I needed to do more. It's actually quite nice to use it essentially as an assistant to basically be like you've given me this."*

The Copywriter practically determines the level of sophistication required for writing, and how to turn that into a prompt. He does this by drawing on his experience and workplace knowledge. He interpretively matches the tools ability to assess the level of tone sophistication, and subsequently the role that the tool plays in his workplace configuration. He then adjusts where it fits in the division of labor. In other words he's able to determine that the appropriate role is an assistant who hasn't demonstrated that it has the intuition to assess writing style. Thus he places it in a role based on his ideal typical notion of an assistant. He understands his responsibility to repair generative AI output in this new configuration, drawing on his commonsense knowledge of his role and skills within that setting.

His prompt interactions are examples of reorientation of respecification of work. Generative AI's contribution in this division of labor is not stable, it is iteratively manipulated to align with his standards. He renders it as a junior assistant, whose outputs must be made legitimate for the campaign he is working on, and its persona is reflexively adjusted in order to do good work.

Thus we see how creative workers must do the work of sustaining workplace order. They do this by drawing on their workplace knowledge, domain expertise and tacit understanding of what is "good enough", and how those tools can play an appropriate role in that workplace.

# 7   STAKEHOLDER IMPRESSION MANAGEMENT, AUTOMATION AND DIVISION OF LABOR

## 7.1   Managing Stakeholder Accountability

Prior to the advent of widespread uptake of LLMs and generative AI, Clarke [4] observed how creative advertisers adhere to background expectancies in their design space. In the process of coming up with advertising campaigns, workers reason through statistical

*Beyond Replacement or Augmentation: How Creative Workers Reconfigure Division of Labor with Generative AI*

output by drawing on knowledge local to their workplace. This included channeling expectations of end clients, which ultimately shaped the outcome of advertising targeting, and matching messaging with that targeting. Now, we see similar work being done to accommodate and configure generative AI tools. The difference is generative AI tools combine insights about a given group of people, with the option to confidently produce image and text based outputs about that group. There is no doubt this is a beneficial tool for creatives. However, we see that users still need to grapple with creating the best output, while ensuring the people paying them are sufficiently acknowledged. This has implications for the generation and content of the idea itself, and also how AI is used to get to the outcome.

A Brand Strategist, describes how end clients are traditionally acknowledged when evaluating the quality of creative work, and the decision of whether to share it or not.

*"Whether that's like..a tagline, or, like, you know, a whole yeah. Just campaign…the strategist kind of represents the consumer in the room, and the account person represents, like the client itself. So we're kind of looking into all those different lenses. To be able to make the call of whether or not we are gonna present the work."*

In this case, the account person (Client Manager) reflects on channeling the needs of the client. They do this by responding to their requests directly and literally, but more importantly by assuming and representing their priorities. Previous work has shown they accomplish this through lay sociological methods, channeling client needs into the discussion [5]. It's instructive to compare traditional work with an example of how a creative works directly with a generative AI tool.

With generative AI the worker often combines traditional client acknowledgment with the task of directly manipulating tools. Creative Director 2, describes how he primes the generative AI tool to be a particular person, and toggles back and forth between the consumer and the client. He does this in order to meet the needs of both the consumer and client audiences. He carefully orients to the needs of both, and also the ways in which separate instances of generative AI tools respond to prompts. He does this in order to iteratively build something that satisfies all stakeholders. Turning back to Sacks [32] this work is done through applying membership categorization devices, invoking separate inference laden categories, and applying them when the creative worker decides they are most relevant.

*"I'll prime it with really interesting stuff..who I think the client is…I'll be like, Hey, pretend you are a 35 year old millennial marketing executive at this company…because I know that's the ultimate audience…after that first prompt is given and I kinda get my results..…..So really, I'm always trying to get it…to pretend like it's the audience for something….Sometimes I'll change that audience between the client and the actual consumer…I toggle a lot."*

Creative Director 2 then describes how he iterates on his work by giving the tool instances of feedback. This demonstrates another example of repairing Generative AI tools with judgements of style and tone, but this time it's done through the lens of client stakeholders:

*"Then I'll give that fake person feedback. So I'll be like, Hey, start thinking about this..like you went too. Trendy. This is too like, try hard. Gen. Z. Can you make this a little bit more professional…or the opposite. This is feeling really dry. Can you inject it with a little bit more? Gen. Z. Personality? So I start kind of going back and forth like it's a writer on my team, like a junior writer who is just brute force creating a bunch of copy for me, and I'm giving them little prompts just to see what it unlocks copy wise… That's where I get in, and I start combining things. I start rewriting things. I like this phrase over here. This word is cool."*

Without saying it directly, the creative worker prompts the generative AI tool to invoke the voice of the client "the ultimate audience" and does this in order to make them accountable [12], or reasonably recognizable to the client. That accountability is realized by the "sound" produced by both mimicking the characteristics of the client (millennial marketing executive), but also adjusting its level of trendiness to address the end consumer (Gen Z). Thus, the indexical expression of "sound more Gen Z" relies on several layers of local workplace context and knowledge. This also demonstrates how the creative director is orienting to an egological principle of organisation performing his own task in individually, but "attentive to the work of others in order to organise the flow of work in a coherent way" [29].

Lastly, Creative Director 2 describes this in light of end consumer and client incentives and reconciling seemingly divergent needs. We see this translate to what he is doing in the tool directly.

*"because to go back to the incentive conversation that we were having…those two things are always intention, because what the consumer wants and what the client wants are actually not the same thing."*

Thus, we see more examples of how the ultimate outcome for this exercise is not simply some objective end state produced solely by the generative AI tool, it's bound up in creative workers ability to artfully combine the expected ego needs of the client and reconcile a perceived mismatch between what the client thinks the consumer needs, and their ability to effectively communicate with that consumer. Here the generative AI tool (in place of a junior copywriter/intern) is used as an efficient vessel. A vessel that is not only highly productive, but flexibly used to impart the expectations and needs of the client and what they want from the "persona" they envision targeting.

These needs extend beyond balancing the audience objectives of the client and consumer. Creatives can be seen orienting to specific departmental and functional needs from their clients. Next, we see an example of "seeing" the needs of a legal department, and shaping the prompt of the LLM accordingly. While working on a promotional case

*Beyond Replacement or Augmentation: How Creative Workers Reconfigure Division of Labor with Generative AI*

study in ChatGPT the Copywriter describes a case study he worked on using ChatGPT. While writing in a style of the client's brand, and targeting a specific consumer segment, he found that he struggled through his prompts and revisions. He found himself constantly orienting to the legal department.

*"Like I did one with somebody from [grocery retail chain] the other day, and it was super difficult..They can't speak to literally anything in the company. You know what I mean like…this will not pass our legal. If if I say literally anything so kind of like roundabout ways, and I'm like you are making what like…Difficult. Yeah."*

In this case, it's difficult for the Copywriter to consider the legal department, while expecting useful output from the tool. He describes how he has to shape the finished product and share back with the client after generating the case study. This ultimately impacts the form the output takes. In reflecting how he oriented to the legal department as a scenic feature in shaping the output, he also reflexively makes his response to that constraint accountable to that legal collaborator through removal of language and details that may seem controversial. This could not be accomplished by technology alone.

In a similar sense, we can see how several creatives adhere to shifting levels of client stigma and comfort with how AI is used in the creative outcome:

The Copywriter describes disclosure rules: *"like you don't have to disclose. If it's in a pitch. There's no actual thing that you're selling. Anyway, you don't have to disclose it, but once it's actually being done, you have to disclose that it, then an AI is is being used."*

A Creative Producer describes legal obligations:

*"But I think that there is like an ethical thing. We have signed a document about using AI for text generation. Right?"*

Executive Creative Director describes the limits of using AI for imagery, and how boundaries are enforced the closer they get to a client:

*"If I'm looking for like specific imagery of things that I think are like relatively easy to understand. Like I need an image of an orange on a blue background that is lit like, you know. Instead of trying to like scour the Internet to find it, I'll prompt it in AI….from a visual perspective, it's never made it to like consumer facing production. I think there's also a little bit of like ethics and AI that comes in from an agency perspective… it's really more like…a behind the scenes thing…a tool amongst ourselves, or proof of concept to clients. We will show AI stuff to clients. We use it a lot in pitches"*

Further, Art Director 1 describes how there is still a "stigma" to using AI which spans both client and internal division of labor concerns. Here "mowing someone's lawn" refers to an uninvited veering into someone else's realm of responsibility and completing their work

for them:

*"There's still stigma to it. I think a lot of people are scared, but they don't want us. They don't want to mow anyone else's lawn."*

He goes on to describe in more detail how these AI disclosures intersect with boundaries setup by themselves to defend the economics and value of their work. Thus, he manages the tension between client AI comfort, economics and producing a good end product.

*"we put together a use policy and a thought piece on how we think you should use AI to every client that we have…getting to the point where client has to tell us not to use AI…it's getting baked into everything now photoshop whatever….some of them are happy with us using it for concepting but don't want AI deliverables..we don't want that either because it goes against our idea of human value add…that conversation happens before any work"*

We see now that creatives also orient to AI stigma and acceptability, which materially impacts the finished product. This makes those additional constraints accountable to the end client. Next, we will see that these increasing levels of client comfort with AI usage, combined with efficiency and economic value opens up new ways of working. These ways of working need to be interpretively managed.

### 7.2 Templatizing Trust for Scalable Output

Creatives showed us how their work reflexively orients to client needs, departmental pressure and attitudes toward AI as non-empirically verifiable "scenic features". The client "figured in the collaborative reasoning" of creatives and "played a role in the reciprocal persuasion through which the internal configuration of the design space was (in part) constructed in media res" [34]. This reflexivity shapes generative AI outputs, as creatives help the tool "see" these constraints and work within them [25]. This is work that could not be accomplished by technical means alone. The interviews point us to some of the lay sociological methods employed in the course of work, and the importance of the everyday practices contextually bound to the creative workplace [12]. Technical and non technical resources are artfully arranged to accomplish these creative outcomes.

Once creatives build up to an outcome they are happy with, reflecting the needs of the client, they build replication tools and practices into their workflow. They craft a division of labor that helps them replicate work originally accomplished one-off. Practices endemic to this replication work are geared toward resolving and making accountable the dual concern of:

1. Workload demands on the creative agency: eliminating part of the workflow of creating custom creative for a similar persona.

*Beyond Replacement or Augmentation: How Creative Workers Reconfigure Division of Labor with Generative AI*

2. Orienting to the dual challenge of communicating clearly to clients while alleviating client anxieties about "safeguarding their brands"

Once the output is finalized, some creatives figure out how to scale "personas" on behalf of the client. They do this in ways that allow them to turn around work in a templatized form that wouldn't have been possible without AI. In these templates, personas can take on a very specific form and modality. Looking back at how creatives orient to client attitudes toward AI, it's clear that the path to scaling and templatizing outcomes requires acknowledgement of client comfort with AI, and how it's used on their behalf. Here we see how one agency has replicated the audible characteristics of voices of brands ("voice talent") that play in their retail locations. This allows the agency to easily produce audio work for clients when new work arises.

Creative Director 3 describes how they created a reusable voice persona for large retail store clients:

*"So we've actually replicated the voices of all of our retail stores. Now in through AI so we've captured very…close to the voice of, say, a [major retailer] or a [grocery chain] and that whole process now takes about an hour….whenever you hear an internal message, when you're walking through a grocery store or a pharmacy, [agency] has voice talent that does that on behalf of brand…a lot of brands choose the option to use the familiar voice of the store for the creative, because it lends credibility to that. [The] ad doesn't feel like an ad. It feels more like a we got to say a well timed suggestion or mention, yeah, but that voice talents. Usually one person."*

He goes on to describe how this reduces both time and effort to produce a new voice ad/suggestion:

*"So we basically just kind of created very similar voice types to the voices of the stores that we now can offer brands. And again, since there's no human required right? It's just a lot quicker…We make sure there's pauses and breaths, and all that…So we've just, you know, it started out as a as an experiment about a year and a half…ago, and now, I like, I said, 50% of our work is done this way.. driving efficiencies with creative production and things like that…it's taken a 14 to 21 day turnaround, and it's turned it into basically like 3 days, maybe 2..make sure we've got the voice of the brand."*

AI also helps creatives navigate the tension of clients safeguarding their brand as intellectual property. Some clients are concerned that by entrusting their brands and personas with others, they may be misinterpreted. The communicative capabilities of generative AI image generation helps assuage these concerns in preliminary stages, before final designs are delivered.

Art Director 2 describes how she uses imagery to help brands see higher fidelity/literal design:

*"But I think clients…they want to see exactly what you're thinking and like this very initial stage… I think it definitely helped with sort of having people to imagine what we're imagining…I kind of wish we had like a little wire from brain to brain, and I think as…But I'm wondering like, will AI be able to replace our brain and just like fully take over that…."*

She also goes on to describe the concerns clients have about safeguarding their brand, and how AI is a communicative tool to alleviate those concerns.

*"clients when they reach out to agencies there's always like this little concern in their head, of like, I'm passing off my brand to these people who aren't part of our brand, but are now trying to hop onto it and help us with this task….having the aid of AI to like…. really show exactly what we're thinking and like shows them the process of like how things would come to life. I think it gives them more trust"*

The art director acknowledges the need for sharing the preliminary generated image, demonstrated by Garfinkel's [12] documentary method of interpretation, to provide evidence of structured creative underpinnings that respects their brand, and directly acknowledging the interpretive work that happens in order to provide that evidence to the client. Generative AI assists them in doing that, but her desire for "brain to brain" connection is acknowledgement that this interpretive work is required despite the scaling/automating power of AI.

We also heard an example of how a creative agency, without directly referencing it, orients to interpretive outcomes as a repeatable/portable set of recipes. These recipes enable the client the ability to do work with AI directly, on their own terms and under their direct control. An Executive Creative Director describes how his team carefully crafts prompts, which can travel into a client organization with ease. This enables clients to create their own imagery with the generative AI tool. This avoids costly product and brand photo shoots for both the agency and client.

*"For one of our clients like we have, we were…designing a brand for a consumer product and..the images…The client didn't want to have to do [photo] shoots every month for their products. So we created a set of seed images along with the prompts and a little action within Firefly and Adobe Photoshop too as part of their brand kit…their creatives could use those pieces to create new images on a monthly basis. Kind of thing…It's just kind of like that little recipe"*

We can see how this work helps resolve fears that brands have about passing off their Intellectual Property (their brand) to an untrusted party. Generative AI tools can be a way to share it, and then technology and humans at the agency enable them to have a toolkit to bring it back in-house. This is a recent change in the workplace ecology.

Lastly, Creative Director 1 describes how these recipes/templates can also help the agency within their own company, efficiently arrive at outcomes with purpose built GPTs.

*Beyond Replacement or Augmentation: How Creative Workers Reconfigure Division of Labor with Generative AI*

*"I'm I would be open to that and definitely open and can definitely see it as a..tool. Like…I definitely think that the agency itself can have its own GPT that we kind of call to. And I also believe that…I think we could definitely live as separate channels for different tasks."*

Automation is the promise of AI, but the character and organizational position and role of these tools are practically accomplished with commonsense methods tied to the context of the particular industry, workplace and sensibilities/sensitivities of clients. The division of labor between the creative worker, generative AI and the end client is artfully constructed by creative workers. When creating templates, they are reorganizing workflow through automation, but still reflexively manage practical client pressures (brand safety) and doing the work to make them accountable to clients. In other words, the technology automates some things, but not the work to make that templated technology work.

## 4  DISCUSSION

This study contributes to CSCW by demonstrating how working with Generative AI tools involves reflexively respecifying a division of labor. This is more than just a matter of interface design improvements, optimized outputs, adoption rates or tool perception. We show how AI is a part of context bound, situated management of an evolving division of labor. Adhering to an ethnomethodological perspective we describe how creative workers shape generative AI outputs and their relationship with surrounding it according to internal and external organizational assumptions and constraints. This points to new opportunities for managing those organizations and designing systems that account for this interpretive work.

In this case study we see how creative professionals reflexively configure generative AI as a workplace collaborator. This work cannot be accomplished by technology alone, thus the augmentation and replacement arguments are insufficient to fully prepare us for AI in the workplace. Creatives attend to generative AI and non AI collaborators through an egological viewpoint, contextually bound to their creative workplace. They use practical reasoning to repair and adjust generative AIs role, establish boundaries between it, themselves and stakeholders within and outside of their direct organization. They manage the impressions of these human collaborators in generative AI's output, by shaping input in a way that makes it accountable to those stakeholders and specific departments. Even in the most "automated" sense of generative AI usage (creating templates to realize efficiency gains) creative workers must make accountable the brand and AI usage concerns of the recipient.

Now where does this leave us? It's clear to us and others before us, generalized and abstract claims about augmentation vs replacement don't give us enough to go on in order to chart our path forward with generative AI. The success of a system depends on enabling, not restricting workers practical know-how and attention to these practices is required to properly integrate the tools into the flow of creative work.

We must treat generative AI as situated. The divisions of labor and role boundaries contained in that workplace are continuously being constructed, respecified, and their boundaries repaired (including with generative AI itself). And as Button and Sharrock [3] say, "people know they are in a division of labor and what they do has consequences for other people…and you will find that they are working in a particular way because that way they will make the work of someone else easier to do" (p.81). This is true for both aligning the generative AI to the most successful path for the task at hand, and in how they consider the work of other departments, collaborators and end clients, who must also do work themselves.

Efficiency is accomplished locally, and cannot simply be ordained on to organizations with grand promises. There is work to do to make the technology work in the workplaces we present it to. This accomplishment is not the abstract version of how AI will replace workers and/or make business more efficient, it's local to the work at hand. Even with the production of templates, efficiency is itself a situated accomplishment [37]. In that vein, templatized work is not simply automating the output of a series of prompts, it requires context and fit to local workplace settings. As Garfinkel [12] says the "'Irredeemable incompleteness of instructions", like that of a templated prompt for a client, presents a space between those instructions and how they are ultimately implemented, in this case how the tool is used and accepted by both the agency and client [29]. Some of that interpretive work is shown here, when considering the efficiency needs of the agency, while orienting to turnaround time expected, and brand needs of the client. There are likely many more of these examples.

If we look beyond simplistic replacement, augmentation and efficiency narratives we see more nuanced ways to approach generative AI technology. We see this both in terms of the form that it takes to support the workplace practices, but also how organizations specifically implement AI, and manage their workforce in relation to it:

1. *System design should reflect an egological viewpoint. We should design generative AI systems that afford organizations and workers the opportunity to intricately assign and shape roles, and make stakeholder needs more visible in their outputs.* Templates and associated collaborative work should be looked at as design opportunities. For example, creative agencies and end customers could agree upon a set of non-negotiables for a brand (E.g. must be a friendly and optimistic "brand voice", must be more professional when posting on Linkedin, fun when on Tiktok, avoid humor for a financial services brand, include diverse representation). These non-negotiables could be included in the pre-training and fine-tuning, and "auto evaluated" (AI automatically assessing the output of AI). The results could be shared back to both the creative agency and client to have a transparent conversation of how far the creative has veered outside of the boundaries that the brand is comfortable with (including the specific boundaries from the legal department we saw earlier). This is a "control room" of sorts where agency and client leadership can have a finger on the pulse of where the last mile of creative work is taking them, allowing for customization

    and taste without fully giving up control or scale of sharing those repeatable creative recipes [17]. We've seen the value of constraints in the creative process [5], and providing this transparency to the parties involved early on may actually increase workers ability to be creative and avoid problems such as the legal scenario shared earlier: *"They can't speak to literally anything in the company…if I say literally anything so kind of like roundabout ways".* While we certainly won't negate the necessity of taking an egological viewpoint between a creative agency and end client, nor create the *"little wire from brain to brain"*, if we make that work accountable in more concrete ways, we might free up attention to additional interpretive, collaborative work.

2. *Organizational leaders should look beyond conventional roles of creative workers when deploying both generative AI tools and their staff. They should pay close attention to the interpretive work required to shape, evaluate and reconfigure generative AI.* Managers should not be seduced by false hopes of full automation and templates, and instead look at templatization of things like voice talent, and product imagery less as "problem solved" but as an opportunity to apply the skill of making those automations fit well with expectations of stakeholders. They must carefully recruit, hire and nurture skills in this changing space. The skills required for creative work may not always be found within clean boundaries of current day creative roles (eg. junior vs not). However, as reported elsewhere, we see here that casting generative AI as a junior assistant has some merit. Unfortunately, more senior creative workers obtain refined "assessment" skills on the job, where taste, style and developing empathy for senior stakeholders is learned by doing. Organizational decision makers will still need to nurture and source this talent in creative ways.

Future research in the creative space should involve more stakeholders, as this work is constrained to the creative agency perspective. Further, understanding how work, and divisions of labor unfold in a more diverse set of workplaces would increase understanding of how to organize work in this new era of AI.

Given the rapid pace of adoption and improvement of generative AI, we have a closing window to understand how to organize work and influence the form those tools take, and we should seize these opportunities. Additional empirical insight like the case provided here is not simply theoretical. It can aid practitioners and scholars to better understand the ways in which workplace usage of generative AI and resultant organizational consequences are not technologically deterministic. Instead, they shape and are shaped by members' everyday methods of making it work for the organizations in question.

**APPENDIX**

| Demographic | Sample |
| --- | --- |

| | | |
|---|---|---|
| Gender | Female | 7 |
| | Male | 10 |
| | Non-binary | 0 |
| | No answer | 0 |
| Age | 20-24 | 2 |
| | 25-29 | 5 |
| | 30-35 | 5 |
| | 40-45 | 3 |
| | 45-50 | 2 |
| Role Type | Strategy | 2 |
| | Account Management | 2 |
| | Creative Direction | 6 |
| | Art Direction | 5 |
| | Creative Producer | 2 |

*Beyond Replacement or Augmentation: How Creative Workers Reconfigure Division of Labor with Generative AI*